\documentclass[twocolumn,aps,prl,amsmath,amssymb,showpacs,superscriptaddress]{revtex4-1} \usepackage[latin9]{inputenc} \usepackage{graphicx}
 
\begin{document} \title{Reversible tuning of the surface state in a psuedo-binary Bi$_{2}$(Te-Se)$_{3}$
topological insulator} 

\author{Rui Jiang}

\affiliation{Division of Materials Science and Engineering, Ames Laboratory, Ames, IA 50011, USA} 
\affiliation{Department of Physics and
Astronomy, Iowa State University, Ames, IA 50011, USA} \author{Lin-Lin Wang}

\affiliation{Division of Materials Science and Engineering, Ames Laboratory, Ames, IA 50011, USA} 
\author{Mianliang Huang}

\affiliation{Materials and Metallurgical Engineering Department, South Dakota School of Mines, Rapid City, SD 57701} 
\author{R. S. Dhaka}

\affiliation{Division of Materials Science and Engineering, Ames Laboratory, Ames, IA 50011, USA} 
\author{Duane D. Johnson}

\affiliation{Division of Materials Science and Engineering, Ames Laboratory, Ames, IA 50011, USA} 
\author{Thomas A. Lograsso}

\affiliation{Division of Materials Science and Engineering, Ames Laboratory, Ames, IA 50011, USA} 
\author{Adam~Kaminski}
\email{kaminski@ameslab.gov} 
\affiliation{Division of Materials Science and Engineering, Ames Laboratory, Ames, IA 50011, USA}
\affiliation{Department of Physics and Astronomy, Iowa State University, Ames, IA 50011, USA} 
\begin{abstract} 
We use angle-resolved photoemission spectroscopy to study non-trivial surface state in psuedo-binary Bi$_{2}$Se$_{0.6}$Te$_{2.3}$ topological insulator. We show that unlike previously studied binaries, this is an intrinsic topological insulator with conduction bulk band residing well above the chemical potential. Our data indicates that under good vacuum condition there are no significant aging effects for more then two weeks after cleaving. We also demonstrate that shift of the Kramers point at low temperature is caused by UV assisted absorption of molecular hydrogen. Our findings pave the way for applications of these materials in devices and present an easy scheme to tune their properties.
\end{abstract}

\pacs{ 73.20.--r, 73.21.Fg, 79.60.Bm, 85.75.--d, 71.20.--b, 71.10.Pm, 73.20.At, 73.22.Gk}

\maketitle 

The prediction and subsequent discovery of surface state with nontrivial spin structure in topological insulators\cite{Zhang,Hasan1} holds great promise to revolutionize spintronics and quantum computing. Such surface state has very unique properties as each electronic state in the momentum space is occupied by a single electron, unlike in traditional metals where two electrons with opposite spin reside at each momentum point. This leads to unusual spin structure at the chemical potential, where electrons at opposite momentum have also opposite spin. This obviously causes strong suppression of backscattering. 

While several studies employing STM and ARPES techniques indeed show that electrons are immune to back scattering and nonmagnetic impurities\cite{Roushan,SCZhang}, the magnetic impurities, which break time reversal symmetry, can create additional surface states with odd number of Dirac fermions\cite{Hasan2} and form a gap at Kramers point\cite{Shen}. Other studies of effects of magnetic and nonmagnetic impurities on scattering rate\cite{Valla1} revealed that there is little difference between those two kinds of impurities. 

The nontrivial surface state can surprisingly survive exposure to atmospheric pressure\cite{Zhou}. However, the electronic properties of topological insulator were shown to change significantly in vacuum with time\cite{Baumberger,Damascelli,Zhou, Hasan3} with typical timescale of hours or even minutes after cleaving and usually results in formation of 2D electron gas. The delicate nature of the non-trivial surface state presents therefore series of challenges such as long--term stability and tunability before it can be utilized in a new class of devices. 
Beside most popular topological insulators Bi$_{2}$Se$_{3}$ and Bi$_{2}$Te$_{3}$, the recently grown Bi$_{2}$(Te-Se)$_{3}$ materials shows surprisingly large bulk resistivity\cite{Re} and low carrier concentration\cite{Cava}. By doping Sb into this material, the Dirac cone can be tuned form n--type to p--type topological insulator\cite{Tunable}.

In this Letter we demonstrate that under good vacuum condition the electronic structure of the surface state in Bi$_{2}$Se$_{2x}$Te$_{2-2x}$ remains unchanged over a period of two weeks even after temperature cycling. We also show that this material is a ``true'' topological insulator,
with bulk conduction band located well above the chemical potential and its electronic properties can be reversibly tuned by UV--assisted absorption of atomic hydrogen. This presents a simple way to tune the carrier concentration at the surface by adjusting device temperature in a low--pressure hydrogen atmosphere.

Single crystals were grown using proper ratio of high purity metals of bismuth (99.999 \%), selenium (99.999\%) and tellurium (99.999\%) that were sealed in a quartz tube and melted into an ingot in an induction furnace to homogenize the composition. The ingot was then sealed in a quartz tube with a larger diameter and loaded into a Bridgman furnace. A crystal was grown by withdrawing the quartz tube at 1 mm/hr after being heated to $800^{\circ}$C. Chemistry of the sample was determined using electron probe micro-analysis (EPMA).

\begin{figure} 
\includegraphics[width=3.5in]{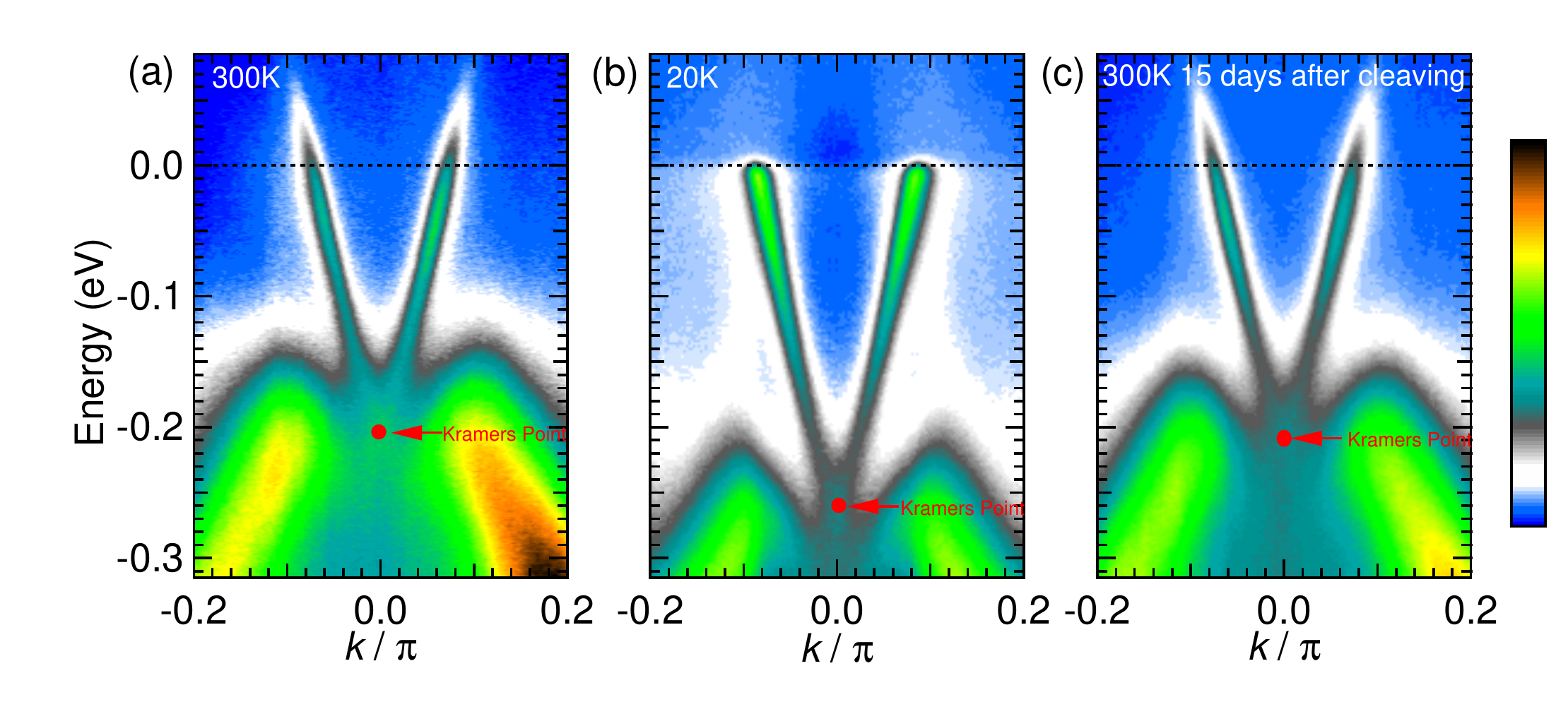}
\caption{\label{fig:300-20-300} 
Long term stability of the Dirac cone and temperature--induced changes in Bi$_{2}$Se$_{2x}$Te$_{2-2x}$. 
(a) Intensity plot along high-symmetry direction at 300K shortly after cleaving. 
(b) The data from the same cleave after cooling to 20K. The band moves to higher binding energy caused by electron doping.
(c) The data from the same cleave as in (a) and (b) at 300~K after 15 days of continuous measurement and temperature cycling,
also showing that the carrier concentration and band position is the same as in freshly cleaved sample.}
\end{figure}

\begin{figure*}\includegraphics[width=7in]{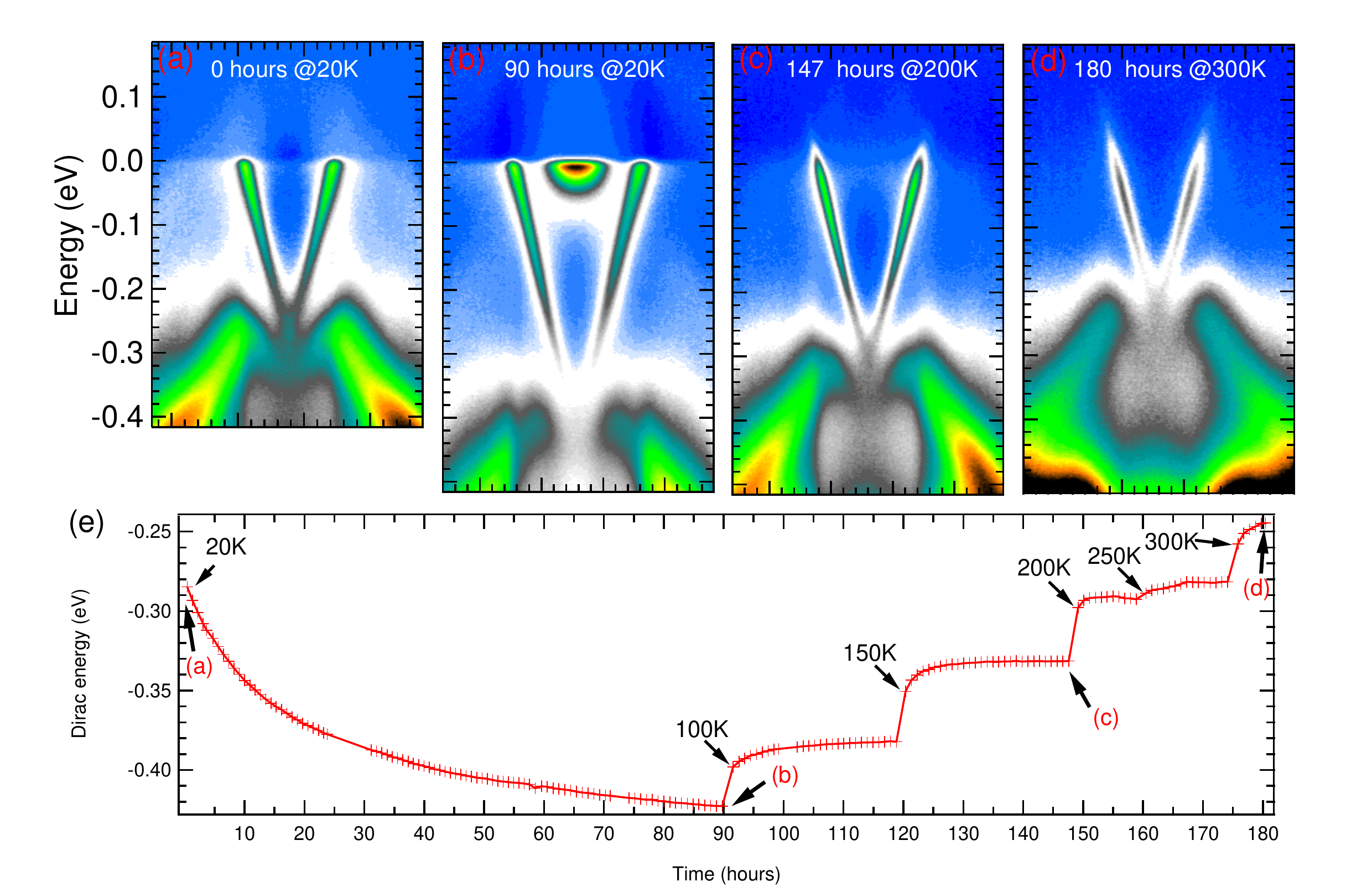}\caption{\label{fig:20-100-150-300} Evolution of the band structure with time and
temperature. 
(a)-(d) Intensity plot at temperatures and time indicated by arrows in panel (e). 
(e) Binding energy of the Kramers point as a function of time upon temperature cycling. Arrows mark the first measurement at a given temperature.} \end{figure*}

ARPES data was acquired using a laboratory-based system consisting of a Scienta SES2002 electron analyzer and GammaData Helium UV lamp.
Samples were cleaved in--situ at room temperature with base pressure in the vacuum system at 5$\times$10$^{-11}$ Tr. All data were acquired
using the HeI line with a photon energy of 21.2 eV. The angular resolution was $0.13^{\circ}$ along and $\sim0.5^{\circ}$ perpendicular to
the direction of the analyzer slits. The energy resolution was set at $\sim6$~meV. Custom designed refocusing optics enabled us to accumulate high statistics
spectra in a short time to study sample aging effects. The results were reproduced on several samples and temperature cycling.

{{} All DFT calculations have been done using VASP\cite{VASP} on the (1$\times$1) surface unit cell for Bi$_{2}$Te$_{3}$(0001) with a slab
of five atomic layers and 12 \AA{}\ of vacuum. The bottom two layers are fixed at bulk positions and the top three layers are free to relax
until the absolute magnitude of force on each atom is reduced below 0.02 eV/\AA{}. A $k$-point mesh of 10$\times$10$\times$1 with a Gaussian
smearing of 0.05 eV and a kinetic energy cutoff of 300 eV were used. }

In Figure \ref{fig:300-20-300}, we examine the temperature dependence and stability of surface band in Bi$_{2}$Se$_{0.6}$Te$_{2.3}$ sample. Intensity plots show almost linear dispersion ofthe surface band and the absence of conduction bulk band at 300~K and 20~K. The Kramers
point, i.~e. the apex of Dirac cone, is below Fermi surface, which indicates a n-type topological insulator. After cooling to 20~K (panel b) the band moves to higher binding energy. This behavior was previously attributed to phonon effects \cite{Valla} or photovoltaic effect
\cite{photovoltaic}. Remarkably, after the sample was warmed up back to 300~K, the band structure recovers to its original state even though
its surface was kept for 15 days in vacuum and exposed to UV and extensive temperature cycling (panel c).

We now focus on the cause of the changes in the band structure at low temperatures. In a fresh cleaved sample, the energy of Kramers point
is -0.22~eV at 300~K, as shown in Figure \ref{fig:300-20-300}a. We performed large number of consecutive measurements for each of the sample
temperatures, with results shown in Figure \ref{fig:20-100-150-300}. In panels a-d, we plot the Angle Resolved Photoemission Spectroscopy
(ARPES) intensity at various temperatures and exposure times. After the sample was kept at 20~K for 90 hours we can observe the bulk
conduction band, which for a clean surface, demonstrates that the chemical potential is located within the bulk band gap. In panel e, we
show the evolution of the binding energy of the Kramers point with time and temperature. Each data point represents separate measurement and
arrows mark the increase in sample temperature. When sample was kept at 20~K, the Kramers pointmoves to higher binding energies, consistent
with electron doping of the surface state. The process is relatively slow, which excludes scenarios involving phonons \cite{Valla} or
photovoltaic effect \cite{photovoltaic}. The associated large time constant suggests chemical doping of the surface as the origin. We note
that the changes significantly slow down with time, indicating saturation effects which are most likely caused by the reduction of the
sticking coefficient with increased coverage. The reverse effect, with a similar time constant, occurs upon warming up. Here, we rapidly
increase the sample temperature to values indicated by the arrows and continuously perform multiple consecutive measurements at each
temperature. The carrier concentration decreases with increasing temperature. In each case, a saturation level is reached when the sample is
kept at fixed temperature for sufficiently long time.

\begin{figure} \includegraphics[width=3.5in]{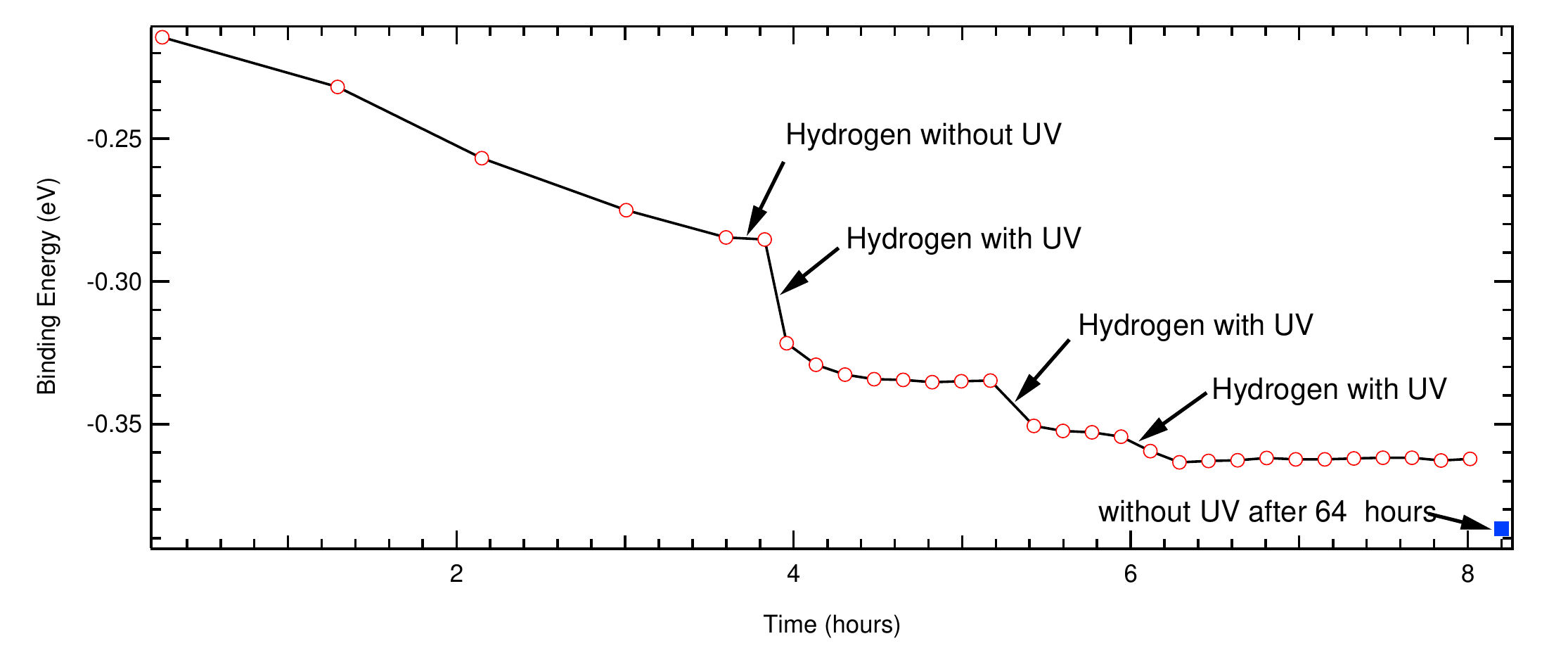}\caption{\label{fig:hydrogen UV} The effects of hydrogen and UV Exposure. All
measurement preformed with sample at 20K. The black arrows indicate brief increase of hydrogen pressure (1$\times$10$^{-7}$ Tr for 10 sec.).
The blue circle indicates binding energy of the Kramers point after 64 hours without UV light.} \end{figure}

The most likely suspect for the change of carrier concentration is the absorption of hydrogen. Even in the best vacuum system made of
stainless steel, hydrogen is omnipresent. Upon cooling, it can condense onto the surface and donate electrons to the surface state. The
binding energy of molecular hydrogen is quite low ($\sim$50 meV). It is possible, however, that UV light used for ARPES is causing its
dissociation at the surface to atomic hydrogen that has a much larger binding energy. To validate our assertion about the origin of the
shift, we dosed small amount of hydrogen into our vacuum system with and without UV. The binding energy of the Kramers point during this
process is shown in Figure \ref{fig:hydrogen UV}. At the beginning of measurement, without additional hydrogen, the binding energy of the
Kramers point increases as in previous case and shifts downward by about 70~meV. We inject hydrogen with UV light switched off for 10
seconds at $10^{-7}$ Torr at time marked by black arrow in Figure \ref{fig:hydrogen UV}. We do not observe any significant change after
injection. Then we inject the same amount of hydrogen in the presence of UV light. Under those conditions the band shifts in energy by
~36~meV, as measured immediately after injection. We repeat this process two more times. The drop is obvious each time, but with decreasing
magnitude, indicating saturation of hydrogen on the surface of sample.

\begin{figure} \includegraphics[width=3.5in]{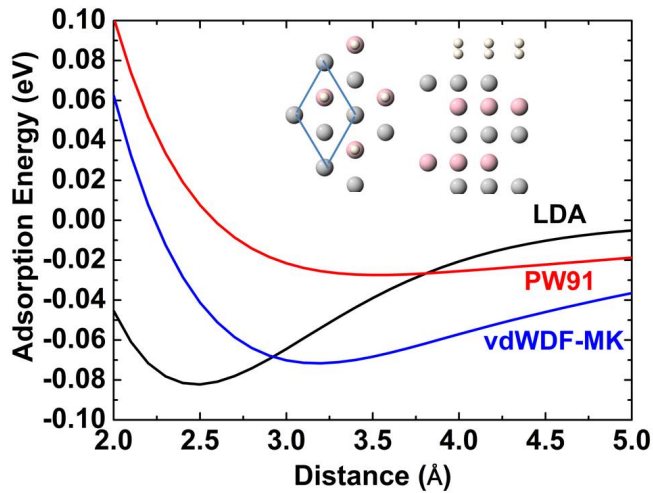}\caption{\label{fig:DFT} Adsorption energy of H$_{2}$ on Bi$_{2}$Te$_{3}$(0001) as a
function of distance to the surface with LDA, PW91and vdWDF-MK as exchange-correlation functional. The inset shows the top and side views of
the relaxed structure. The (1$\times$1) surface unit cell is highlighted in the top view. Red, gray and white spheres stand for Bi, Te and
H, respectively. } \end{figure}

To support our proposition that the origin of electron doping and downward shift of Dirac point is due to absorption of atomic H, we used
density functional theory (DFT)\cite{HK,KS} to calculate the adsorption of H$_{2}$ and H on Bi$_{2}$Te$_{3}$(0001) surface. Figure
\ref{fig:DFT} shows the DFT adsorption energy of H$_{2}$ as a function of distance to the surface with different exchange-correlation
functionals at the hcp site with an out-of-plane orientation for H$_{2}$. The data clearly shows that the interaction between H$_{2}$ and
Bi$_{2}$Te$_{3}$(0001) surface is of van der Waals type. PW91\cite{PW91} gives a very weak binding of $-$28 meV at 3.7 \AA{}\ and
LDA\cite{LDAPZ} gives a stronger binding of $-$80 meV at 2.4 \AA{}, which is closer to $-$71 meV at 3.2 \AA{}\ from the more accurate
description of the system by the van der Waals exchange-correlation functional\cite{vdWMK}. Upon full relaxation, the bond length of the
adsorbed H$_{2}$ is 0.77 \AA{}, only slightly longer than the 0.75 \AA{}\ of the free H$_{2}$ molecule. The relaxation of the surface atoms
is negligible. The adsorption energy at the three adsorption (fcc, BRIDGE (brg) and top) sites is 5, 11 and 36 meV higher than the hcp site,
respectively. The difference in adsorption energy on the same site with different orientations of H$_{2}$ is less than 5 meV.

In contrast, the interaction between atomic H and Bi$_{2}$Te$_{3}$(0001) surface is much stronger, with a binding energy of $-$1.41 eV at
the brg site (in reference to a free atomic H), followed by $-$1.08, $-$0.99 and $-$0.92 eV at the top, fcc and hcp sites, respectively. The
space among the surface atoms can accommodate atomic H very well, the adsorbed H is in a co-planar position to surface Te atoms at all
sites, except the top site, giving a binding distance of 1.73 \AA{}. Although the adsorption energy in reference to a free H$_{2}$ molecule
is1.02 eV, which means that H$_{2}$ does not dissociate on Bi$_{2}$Te$_{3}$(0001) surface, the presence of UV light during ARPES measurement
can produce atomic H, as confirmed experimentally. In supporting evidence, we also directly calculated the shift of Dirac point in the
surface band with different H$_{2}$ coverage. The downward shift due to the weak H$_{2}$-surface interaction is about 20 meV, too small
compared to the shift of 100 meV observed in experiment.

We have carefully detailed that the topological insulator behavior of Bi$_{2}$Se$_{0.6}$Te$_{2.3}$ remains unchanged and/or controllable
over extended periods under good vacuum conditions,distinct from commonly observed Rashba effects for example. We also showed that the Dirac
cone electronic properties can be reversibly tuned by UV--assisted adsorption of atomichydrogen, where DFT calculations confirm the assisted
energetics and adsorption characteristics.

Research supported by the U.S. Department of Energy, Office of Basic Energy Sciences, Division of Materials Sciences and Engineering. Ames
Laboratory is operated for the U.S. Department of Energy by Iowa State University under Contract No. DE-AC02-07CH11358.

\begin{thebibliography}{10}

\bibitem{Zhang} B. A. Bernevig, T. L. Hughes, \& S.-C. Zhang, Science 314, 1757 (2006).
\bibitem{Hasan1} D. Hsieh $et\ al$., Nature 452, 970 (2008).

\bibitem{Roushan} P. Roushan $et\ al$., Nature 460, 1106 (2009).

\bibitem{SCZhang} X.-L. Qi and S.-C. Zhang, Rev. Mod. Phys. 83, 1057 (2011).

\bibitem{Hasan2} L. A. Wray $et\ al$., Nature Phys. 7, 32 (2010).

\bibitem{Hasan3} D. Hsieh $et\ al$., Phys. Rev. Lett. 103, 146401 (2009). 
\bibitem{Shen} Y. L. Chen $et\ al$., Science 329, 659 (2010). 
\bibitem{Valla1} T. Valla $et\ al$., Phys. Rev. Lett. 108, 117601 (2012).

\bibitem{Damascelli} Z. H. Zhu $et\ al$., Phys. Rev. Lett. 107, 186405 (2011). 
\bibitem{Baumberger} P. D. C. King $et\ al$., Phys. Rev. Lett. 107, 096802 (2011). 
\bibitem{Zhou} C. Chen $et\ al$., Proc. Natl Acad Sci. USA 109, 3694 (2012).

\bibitem{Re} Zhi Ren $et\ al$., Phys. Rev. B 82, 241306(R) (2010).
\bibitem{Cava} Shang Jia $et\ al$., arxiv 1112, 1648v1 (2011).
\bibitem{Tunable}T. Arakane $et\ al$., Nat. commmun. 3, 636 (2012).

\bibitem{Valla} Z. H. Pan $et\ al$., arxiv 1109, 3638 (2011).

\bibitem{photovoltaic} S. R. Park $et\ al$., New J. Phys. 13, 013008 (2011). 
\bibitem{HK} P. Hohenberg and W. Kohn, Phys. Rev. 136, B864 (1964). 
\bibitem{KS} W. Kohn and L. J. Sham, Phys. Rev. 140, A1133 (1965). 
\bibitem{VASP} G. Kresse and J. Furthm�ller, Phys. Rev. B 54, 11169 (1996). 
\bibitem{LDAPZ} J. P. Perdew and A. Zunger, Phys. Rev. B 23, 5048 (1981).
 \bibitem{PW91} J. P. Perdew and Y. Wang, Phys. Rev. B 45, 13244 (1992). 
\bibitem{vdWMK} J. Klimes, D. R. Bowler and A. Michaelides, Phys. Rev. B 83, 195131 (2011).
\end{thebibliography}
 \end{document}